\documentstyle[amsmath,11pt]{article}

\begin{document}

\date{\empty}

\title{{\bf Averaging anisotropic cosmologies}}

\author{John D. Barrow${}^1$\thanks{email:
j.d.barrow@damtp.cam.ac.uk} and Christos G.
Tsagas${}^2$\thanks{email: tsagas@astro.auth.gr}\\
{\small ${}^1$DAMTP, Centre for Mathematical Sciences, University of
Cambridge}\\ {\small Wilberforce Road, Cambridge CB3 0WA, UK}\\
${}^2${\small Section of Astrophysics, Astronomy and Mechanics,
Department of Physics}\\ {\small Aristotle University of
Thessaloniki, Thessaloniki 54124, Greece}}

\maketitle

\begin{abstract}
We examine the effects of spatial inhomogeneities on irrotational
anisotropic cosmologies by looking at the average properties of
anisotropic pressure-free models. Adopting the Buchert scheme, we
recast the averaged scalar equations in Bianchi-type form and close
the standard system by introducing a propagation formula for the
average shear magnitude. We then investigate the evolution of
anisotropic average vacuum models and those filled with pressureless
matter. In the latter case we show that the backreaction effects can
modify the familiar Kasner-like singularity and potentially remove
Mixmaster-type oscillations. The presence of nonzero average shear
in our equations also allows us to examine the constraints that a
phase of backreaction-driven accelerated expansion might put on the
anisotropy of the averaged domain. We close by assessing the status
of these and other attempts to define and calculate `average'
spacetime behaviour in general relativity.\newline\newline PACS
Numbers: 04.20.-q, 04.40.-b, 98.80.-k, 98.80.Jk
\end{abstract}

\section{Introduction}
The question of whether an inhomogeneous model of the universe
behaves on the average like a homogeneous solution of Einstein's
equations is a long-standing problem, directly related to the
question of how to average general-relativistic
spacetimes~\cite{SF}-\cite{ES}. The problem lies in the generic
non-linearity of general relativity and in the difficulty of
establishing a unique and unambiguous way of averaging the spacetime
metric without throwing away crucial information in the averaging
process. A number of cosmologists have argued that the averaging
problem may be central to our understanding of the recent expansion
history of the visible universe. When it was suggested, recently,
that structure formation could be responsible for the observed
acceleration of the universe~\cite{R1}-\cite{KMNR2}, averaging
techniques were employed to study the dark energy problem in terms
of kinematic backreaction effects from spatial
inhomogeneities~\cite{B1,R2}. Although the idea of backreaction has
been criticized, primarily on the basis of current
observations~\cite{GCA}-\cite{IW}, it is attractive in principle
because it can solve the dark energy and the coincidence problems
without appealing to a cosmological constant, speculative
quintessence field, or non-linear corrections in Einstein's theory
of gravity.

General relativity has been applied to a range of averaging scales
depending on the nature of the physical system under study.
Cosmology applies to the largest astronomical scales, which
typically extend over a significant fraction of the Hubble radius.
Nevertheless, the major issue of defining a suitable averaging
process remains open. When employing averaging techniques to study
the backreaction of cosmological perturbations, one usually averages
scalars like the density or the volume expansion rate. These become
the scale-dependent parameters that are compared with observations.
Averaging such scalars over a Riemannian spacelike region gives the
effective dynamical sources that an observer would expect to
measure. Thus, the averaged equations isolate an explicit source
term that is commonly referred to as the `backreaction' term. The
latter quantifies the deviation of the average model from a given
`fitting' model, which is usually the Friedmann-Robertson-Walker
(FRW) universe, though not always~\cite{PS}.

Here, we do not use a Friedmannian universe, but an anisotropic
Bianchi-type cosmology that contains the FRW models as subcases. We
consider vacuum, or dust-dominated, irrotational Bianchi models with
zero or isotropic local spatial curvature, aiming to describe the
effective evolution of their averaged counterparts. We do this by
writing the standard average formulae in Bianchi form (i.e.~with
explicit shear terms in the left-hand side) and by closing the
system with a propagation equation for the mean shear magnitude.
Within this Bianchi-type interpretation of the averaged equations
the backreaction term is expressed as a difference between the
mean-square fluctuations in the expansion and shear scalars. For a
vacuum background with Euclidean 3-spaces, we find that the source
term vanishes and the average equations retain their local form.
When pressureless matter is introduced, however, the backreaction
effects can dominate the early evolution of averaged Bianchi~$I$
models and prevent the familiar Kasner-type singularity from
forming. Assuming axial symmetry, we examine the `bounce conditions'
that avoid the initial singularity for different types and degrees
of anisotropy. We also consider the accelerated expansion of these
models and how this might constrain their average shear anisotropy.

\section{The averaging rule}
The literature contains more than one averaging
method~\cite{I1}-\cite{B4} that has been applied to cosmology
(see~\cite{F} and more recently~\cite{B1,R2}
and~\cite{CPZ}-\cite{BLA}). Here, we will follow the Buchert
approach which builds on the Newtonian theory, where spatial
averaging is relatively simple, before extending to general
relativity by confining to scalar variables~\cite{BE}-\cite{B3}.
Consider a general spacetime filled with irrotational pressureless
matter moving along geodesic worldlines with 4-velocity $u_a$
(i.e.~$u_au^a=-1$ and $\dot{u}_a=u^b\nabla_bu_a=0$). Given a
foliation of flow-orthogonal hypersurfaces, the metric of the
3-dimensional space is $h_{ab}=g_{ab}+u_au_b$. Then, the spatial
averaging of an arbitrary scalar field $\phi=\phi(x^a)$ over a
simply-connected domain ${\cal D}$ is a covariant operation defined
by the averaging operation $\langle\phi\rangle_{\cal D}=\int_{\cal
D}\phi J{\rm d}^3x^{\alpha}/V_{\cal D}$, where the angled brackets
indicate spatial averages, $V_{\cal D}$ is the volume of the domain
in question, and $J=\sqrt{{\rm det}(h_{ab})}$~\cite{BE}. We also
note that Greek indices take the values 1,\,2,\,3 and Latin ones run
from 0 to 3. Following~\cite{B1}, it helps to define an effective
expansion scale factor by $a_{\cal D}=(V_{\cal D}/V_{{\cal
D}_i})^{1/3}$, with $V_{{\cal D}_i}$ representing the volume of the
initial domain. Since the volume $V_{\cal D}$ is generally time
dependent, the average of the time derivative of a locally-defined
scalar differs from the time derivative of the average. This
non-commutativity between spatial averaging and temporal evolution
can be formulated in a simple mathematical rule. For a rest-mass
preserving domain, the commutation between the volume-averaging and
the time-evolution operators leads to~\cite{BE,B3}
\begin{equation}
\langle\phi\rangle_{\cal D}^{\cdot}- \langle\dot{\phi}\rangle_{\cal
D}=\langle\Theta\phi\rangle_{\cal D}-\langle\Theta\rangle_{\cal
D}\langle\phi\rangle_{\cal D}\,,  \label{Bcl}
\end{equation}
where $\Theta$ is the local volume expansion rate between
neighbouring worldlines. This scalar coincides with the trace of the
local expansion tensor and also with that of the extrinsic curvature
tensor. The right-hand side of the above rule generally introduces
source terms to the averaged equations which can be interpreted as
backreaction effects due to the averaging process. Such kinematic
backreaction has recently been considered as a possible solution to
the dark energy problem. We note that the appearance and the nature
of the aforementioned source terms depends on the form of the
averaged equations. Generally, the less information these equations
contain the simpler (if any) the source terms are.

\section{Averaged anisotropic cosmologies}
\subsection{The local evolution}\label{ssLE}
Consider an irrotational, anisotropic, spacetime and introduce a
family of (fundamental) observers living along worldlines tangent to
the future-directed timelike eigenvectors of the Ricci tensor. This
achieves a unique local threading of the spacetime into time and
3-dimensional space (e.g.~see~\cite{EvE}). For a pressure-free
medium, the local shear tensor propagates as~\cite{TM,BT}
\begin{equation}
\dot{\sigma}_{ab}=-\Theta\sigma_{ab}- {\cal R}_{\langle
ab\rangle}\,,  \label{lsab}
\end{equation}
where ${\cal R}_{\langle ab\rangle}$ is the traceless part of the
locally-defined 3-Ricci tensor. The latter describes the geometry of
the spatial hypersurfaces orthogonal to $u_a$ and satisfies the
Gauss-Codacci equation. Hereafter, we will only consider spacetimes
that obey the constraint ${\cal R}_{ab}\sigma^{ab}=0$. For example,
this condition holds when the curvature of the spatial sections
vanishes, or when it is isotropic. Then, the local expansion is
monitored by the following three scalar propagation equations
\begin{equation}
\dot{\rho}+ \Theta\rho=0\,, \hspace{15mm} \dot{\sigma}+
\Theta\sigma=0 \hspace{15mm} {\rm and} \hspace{15mm} \dot{\Theta}+
{\textstyle{\frac{1}{3}}}\Theta^{2}+ {\textstyle{\frac{1}{2}}}\rho+
2\sigma^{2}- \Lambda=0\,,  \label{lpe}
\end{equation}
supplemented by the constraint
\begin{equation}
{\textstyle{\frac{1}{2}}}{\cal R}- \rho+
{\textstyle{\frac{1}{3}}}\Theta^{2}- \sigma^{2}- \Lambda=0\,.
\label{l3Ricci}
\end{equation}
Here, $\rho $ is the matter density, $\Theta $ is the expansion
scalar, $\sigma $ is the shear magnitude, ${\cal R}$ is the Ricci
scalar of the spatial sections, and $\Lambda $ is the cosmological
constant.

We remind the reader that the adopted averaging scheme requires zero
rotation to guarantee the existence of integrable spatial
hypersurfaces. Also, the method has been applied to study the
backreaction effects in pressure-free cosmological environments.
Here, in addition to zero rotation and pressure, we assume that
${\cal R}_{ab}\sigma^{ab}=0$. For all practical purposes the latter
is a very mild constraint and still covers all (non-rotating and
pressure-free) spacetimes with flat spatial sections or isotropic
3-curvature. These include those Bianchi models that contain the FRW
universes as subcases. Primarily, condition ${\cal
R}_{ab}\sigma^{ab}=0$ allows us to reduce Einstein's equations to a
closed set of four scalar formulae. Previous work on the
backreaction effects of spatial averaging did not need a shear
propagation equation and the associated sets of scalar equations
were not closed (in reminiscent of the Newtonian `gravitational
paradox' - see~\cite{BG}).

\subsection{The average evolution}
Because spatial averaging and time evolution do not commute, the
averaged equations contain extra terms, compared to their local
counterparts. By giving to the average formulae the form of a chosen
set of local equations, one can interpret these extra terms as the
backreaction effect of the averaging process. Here, we will write
the averaged formulae in Bianchi-type form by allowing for explicit
shear terms in their left-hand side. Thus, applying law (\ref{Bcl})
to the matter density, the shear magnitude, and the volume
expansion, we may use Eqs.~(\ref{lpe}a)-(\ref{lpe}c) to obtain
\begin{equation}
\langle\rho\rangle^{\cdot}+
\langle\Theta\rangle\langle\rho\rangle=0\,, \hspace{15mm}
\langle\sigma\rangle^{\cdot}+
\langle\Theta\rangle\langle\sigma\rangle=0  \label{ape}
\end{equation}
and
\begin{equation}
\langle\Theta\rangle^{\cdot}+
{\textstyle{\frac{1}{3}}}\langle\Theta\rangle^{2}+
{\textstyle{\frac{1}{2}}}\langle\rho\rangle+
2\langle\sigma\rangle^{2}- \Lambda= {\textstyle{\frac{2}{3}}}\langle
\left(\Theta-\langle\Theta\rangle\right)^{2}\rangle-
2\langle\left(\sigma-\langle\sigma\rangle\right)^{2}\rangle\,,
\label{aRay}
\end{equation}
respectively. Also, the volume-averaged counterpart of
(\ref{l3Ricci}) reads
\begin{equation}
{\textstyle{\frac{1}{2}}}\langle{\cal R}\rangle- \langle\rho\rangle+
{\textstyle{\frac{1}{3}}}\langle\Theta\rangle^{2}-
\langle\sigma\rangle^{2}- \Lambda= -{\textstyle{\frac{1}{3}}}\langle
\left(\Theta-\langle\Theta\rangle\right)^{2}\rangle+
\langle\left(\sigma-\langle\sigma\rangle\right)^{2}\rangle\,,
\label{a3Ricci}
\end{equation}
where the right-hand sides of (\ref{aRay}) and (\ref{a3Ricci})
involve the identity
$\langle\left(\phi-\langle\phi\rangle\right)^{2}\rangle=
\langle\phi^{2}\rangle-\langle\phi\rangle^{2}$. The above set is a
system of four effective Einstein equations for spatially averaged
scalar variables in inhomogeneous irrotational universes with zero
fluid pressure and ${\cal R}_{ab}\sigma^{ab}=0$. Within this
environment Eqs.~(\ref{ape})-(\ref{a3Ricci}) are exact and there is
no need to assume that the inhomogeneity and the anisotropy are
small perturbations~\cite{B1,R2}.

Comparing Eqs.~(\ref{ape}) to (\ref{lpe}a) and (\ref{lpe}b) we
notice that, despite the non-commutativity between volume averaging
and time evolution, the averaged formulae have preserved the form of
their local counterparts. This means that
\begin{equation}
\langle\rho\rangle\propto a_{\cal D}^{-3} \hspace{15mm} {\rm and}
\hspace{15mm} \langle\sigma\rangle\propto a_{\cal D}^{-3}\,,
\label{arho-asigma}
\end{equation}
where $a_{\cal D}$ is the average scale factor smoothed out over the
domain ${\cal D}$ (with
$\dot{a}_{\mathcal{D}}/a_{\mathcal{D}}=\langle\Theta\rangle/3$).
However, the averaging process has led to extra terms in the
right-hand sides of (\ref{aRay}) and (\ref{a3Ricci}), collectively
given by the domain-dependent scalar
\begin{equation}
{\cal Q}\equiv {\textstyle{\frac{2}{3}}}
\langle\left(\Theta-\langle\Theta\rangle\right)^{2}\rangle
-2\langle\left(\sigma-\langle\sigma\rangle\right)^{2}\rangle=
-{\textstyle{\frac{2}{3}}}\left(\langle\Theta\rangle^{2}
-\langle\Theta^{2}\rangle\right)+
2\left(\langle\sigma\rangle^{2}-\langle\sigma^{2}\rangle\right)\,.
\label{cQ}
\end{equation}
We interpret the above as the backreaction effect of spatial
averaging on pressure-free Bianchi~I models, since in its absence
the averaged formulae recover the form of their local counterparts
(compare to Eqs.~(\ref{lpe}), (\ref{l3Ricci})). Note that
$\mathcal{Q}$ can be either positive or negative, and its value is
decided by the amount of kinematic anisotropy. In particular, ${\cal
Q}$ is positive when the shear mean-square fluctuation is less than
a third of that in the volume expansion.

\subsection{The kinematic backreaction term}\label{ssKBRT}
The form of the ${\cal Q}$-term depends on the specifics of the
adopted cosmology. For instance, in empty models with zero
cosmological constant and 3-curvature, the local shear and the
expansion are related by $\Theta/\sqrt{3}=\sigma$ (see
Eq.~(\ref{l3Ricci}) earlier). This immediately ensures that
$\langle\Theta\rangle/\sqrt{3}= \langle\sigma\rangle,$ and
subsequently that
$\langle\Theta\rangle^{2}/3=\langle\sigma\rangle^{2}$. Hence,
following (\ref{cQ}), the backreaction term vanishes identically,
and Eqs.~(\ref{aRay}), (\ref{a3Ricci}) maintain the form of their
local counterparts.

When there is matter or nonzero 3-curvature, the backreaction
effects do not generally vanish. In those cases we can monitor the
dynamical evolution of ${\cal Q}$ via a consistency/integrability
condition. Taking the time derivative of (\ref{a3Ricci}) and then
using (\ref{ape}), (\ref{aRay}) we obtain
\begin{equation}
\dot{{\cal Q}}+ 2\langle\Theta\rangle{\cal Q}= -\langle{\cal
R}\rangle^{.}-
{\textstyle{\frac{2}{3}}}\langle\Theta\rangle\langle{\cal
R}\rangle\,.  \label{cQdot}
\end{equation}
It follows that in the absence of pressure and rotation, and also
for ${\cal R}_{ab}\sigma^{ab}=0$ in our case, the evolution of the
backreaction term depends on that of the average 3-Ricci scalar.
When ${\cal R}\equiv 0$ or $\langle{\cal R}\rangle\propto\langle
a\rangle^{-2}$, for example, we find that
\begin{equation}
{\cal Q}\propto a_{\cal D}^{-6}\,.  \label{cQ1}
\end{equation}
Overall, a positive $\mathcal{Q}$ mimics the dynamics of a dark
energy component and tends to accelerate the expansion, whereas a
negative one acts as an effective dark matter source (see
Eq.~(\ref{aRay})). In particular, a negative $\mathcal{Q}$ behaves
like an effective fluid with a `stiff' equation of state (e.g.~a
free scalar field) or like an additional shear source.

\section{Backreaction effects}
\subsection{Backreaction and Kasner-type singularities}
The scalar $\mathcal{Q}$ carries the collective backreaction effects
from spatial inhomogeneities. This could mean a considerable loss of
information, relative to that contained in the local equations. For
instance, the sign of the backreaction term is crucial but unknown.
There is an advantage, however, because the compactness of
$\mathcal{Q}$ makes it an easier to handle quantity. Thus, a
positive $\mathcal{Q}$ was recently suggested as potential cause of
late-time acceleration in FRW-type average cosmologies. Here we will
consider average Bianchi-type models and examine whether analogous
backreaction effects can modify (or even remove altogether) standard
features of their conventional counterparts.

Integrating (\ref{lpe}a) and (\ref{lpe}b), we find that $\rho
\propto a^{-3}$ and $\sigma^{2}\propto a^{-6}$, with the latter
holding for ${\cal R}_{ab}\sigma^{ab}=0$. This means that shear
anisotropies dominate the early evolution of these models, which
approach the Kasner-type singularities of the vacuum Bianchi~$I$
spacetimes (with $\Lambda=0$). This singular behaviour is not
guaranteed for the averaged counterparts of these cosmologies
because of the backreaction term and particularly because of
(\ref{cQ1}). Given that both $\langle \sigma\rangle^{2}$ and ${\cal
Q}$ propagate as $a_{D}^{-6}$ while $\langle\rho\rangle\propto
a_{D}^{-3}$, the early stages of the averaged kinematics is
dominated by the sum ${\cal Q}-2\langle\sigma\rangle^{2}=
2\langle\left(\Theta-\langle\Theta\rangle\right)^{2}\rangle/3-
2\langle\sigma^{2}\rangle$. When the latter is negative and
$\langle\Theta\rangle<0$ initially, the collapse will proceed
unimpeded and $\langle\Theta\rangle\rightarrow-\infty$ within a
finite amount of time. Otherwise the averaged model should bounce
before reaching the singularity.

Additional information is obtained by expressing ${\cal Q}$ in terms
of the averaged individual expansion rates. Assume, for example and
also for the sake of simplicity, that the inhomogeneous spacetime is
axially symmetric and has zero 3-curvature. Then
$\Theta=2\alpha+\beta$, $\rho=\alpha(\alpha+2\beta)$ and
$\sigma=(\alpha-\beta)/\sqrt{3}$. In this notation
$\alpha=\dot{a}/a$ and $\beta=\dot{b}/b$ are the expansion rates
along the two main axes, with $a$ and $b$ representing the
associated scale factors of the axisymmetric Bianchi~$I$
universe~\cite{LL,Ba1}. Within this environment, expression
(\ref{cQ}) reads
\begin{equation}
{\cal Q}=
2\left(\langle\alpha^{2}\rangle-\langle\alpha\rangle^{2}\right)+
4\left(\langle\alpha\beta\rangle
-\langle\alpha\rangle\langle\beta\rangle\right)\,,  \label{cQ2}
\end{equation}
allowing for ${\cal Q}$ to be positive, negative, or
zero.\footnote{When $\alpha=-2\beta$ expression (\ref{cQ2}) ensures
that ${\cal Q}=0$ and therefore the absence of any averaging
backreaction effects. This is expected since $\alpha=-2\beta$ means
$\rho=0$ and we know that ${\cal Q}=0$ in the vacuum models (see
\S~3.3).} Hence, if the expansion is nearly isotropic (i.e.~for
$\alpha\simeq\beta$), the bounce condition ${\cal
Q}>2\langle\sigma\rangle^{2}$ is satisfied when
$\langle(\alpha-\langle\alpha\rangle)^{2}\rangle>0$. For highly
anisotropic expansion, with $\alpha\gg\beta$, we have
$\langle\sigma\rangle^{2}\simeq\langle a\rangle^{2}/3$. In this
case, we will have a bounce when the mean-square fluctuation of
$\alpha$ is greater than $\langle\alpha\rangle^{2}/3$. Finally, for
$\alpha\ll\beta$, the backreaction effects can prevent the
$\langle\Theta\rangle\rightarrow-\infty$ singularity provided that
$\langle\alpha\beta\rangle
-\langle\alpha\rangle\langle\beta\rangle>\langle\beta\rangle^{2}/6$.

It is well known that the Mixmaster behaviour of the Bianchi~$IX$
models consists of a sequence of chaotically alternating Kasner-like
time intervals where the collapsing and expanding scale factors
interchange (e.g.~see~\cite{LL} and also~\cite{Ba2,CB}). The
possibility that backreaction effects due to spatial averaging could
alleviate the Kasner-type singularities of the Bianchi~$I$ models,
raises the question as to whether the same effects could also modify
the Mixmaster regime of the Bianchi~$IX$ spacetimes. The possible
role of stochastic spatial fluctuations was discussed in~\cite{Ba3}.
It was argued there that the cumulative stochastic effects from
nearby gravity-wave perturbations could transform the local Einstein
equations into stochastic differential equations with significant
damping effects on the chaotic oscillations. Here, we see the
cumulative effect of the stochastic inhomogeneities mimics a scalar
field of the sort that is known to stop the continuation of
Mixmaster oscillations on approach to the initial singularity (if it
exists)~\cite{BK}. Based on these we conjecture that, within the
framework of the Buchert averaging scheme, Mixmaster oscillations
will be unstable at early times.

\subsection{Accelerated expansion and average anisotropy}
Let us suppose that the aforementioned backreaction effects lead to
the acceleration of the averaged domain. The presence of nonzero
mean shear in the equations allows us to consider the constraints
imposed on the average anisotropy of the region. Using the
scale-dependent average scale factor $a_{\cal D}$, defined so that
$\dot{a}_{\cal D}/a_{\cal D}=\langle\Theta\rangle/3$, the averaged
Raychaudhuri equation (see (\ref{aRay})) takes the form
\begin{equation}
3\frac{\ddot{a}_{D}}{a_{D}}=
-{\textstyle{\frac{1}{2}}}\langle\rho\rangle-
2\langle\sigma\rangle^{2}+ \Lambda+ {\cal Q}\,.  \label{aRay1}
\end{equation}
Here, the backreaction term can be interpreted as an effective
energy-density term or an effective shear term.\footnote{One can
introduce a chosen `background' to the average equations by defining
a suitable reference model. Thus, the time-dependent expansion and
shear fields, $H=H(t)$ and $\Sigma_{ab}=\Sigma_{ab}(t)$
respectively, define a homogeneous background into the averaged
formulae~\cite{BE}. Inhomogeneous deviations from these reference
fields are considered by setting $\Theta=H+\hat{\Theta}$ and
$\sigma_{ab}=\Sigma_{ab}+\hat{\sigma}_{ab}$. This type of
decomposition is not essential for our purposes. If adopted,
however, the averaged Raychaudhuri equation (e.g.~see (\ref{aRay1}))
will take the form
\begin{equation}
3\frac{\ddot{a}_{D}}{a_{D}}=
-{\textstyle{\frac{1}{2}}}\langle\rho\rangle+
{\textstyle{2\over3}}\left(\langle\hat{\Theta}^2\rangle
-\langle\hat{\Theta}\rangle^2\right)- 2\Sigma^2-
2\langle\Sigma_{ab}\hat{\sigma}^{ab}\rangle-
2\langle\hat{\sigma}\rangle^{2}+ \Lambda\,,  \label{adRay}
\end{equation}
with an analogous expansion for (\ref{a3Ricci}). Clearly, when the
`background' is isotropic all the $\Sigma$-terms vanish
identically.} In any case, and in the absence of a cosmological
constant, the volume expansion of the averaged domain will
accelerate provided that
\begin{equation}
{\cal Q}> {\textstyle{\frac{1}{2}}}\langle\rho\rangle+
2\langle\sigma\rangle^{2}\,,  \label{acc}
\end{equation}
which here holds for zero or isotropic 3-curvature (strictly
speaking as long as ${\cal R}_{ab}\sigma ^{ab}=0$). We note that if
$\mathcal{Q}$ is negative, namely if the mean square fluctuation in
the shear is greater than a third of the mean square fluctuation in
the volume expansion (see~(\ref{cQ})), the domain will decelerate
instead of accelerating. We have no information on the sign of the
backreaction term however. Because of this ambiguity, early on,
kinematic backreaction was also suggested as an effective
dark-matter source~\cite{B5}.

Putting these matters aside, we will use our Bianchi-based
interpretation of the backreaction effects and the form of the
average equations to constraint the mean anisotropy of the
accelerating region. To proceed we introduce the following
domain-dependent, dimensionless set:
\begin{equation}
\Omega_{\rho}=\frac{3\langle\rho\rangle}{\langle\Theta\rangle^2}\,,
\hspace{10mm} \Omega_{\sigma}=
\frac{3\langle\sigma\rangle^2}{\langle\Theta\rangle^2}\,,
\hspace{10mm} \Omega_{\cal R}= -\frac{3\langle{\cal R}\rangle}
{2\langle\Theta\rangle^2}\,, \hspace{10mm} \Omega_{\cal Q}=
-\frac{3{\cal Q}}{2\langle\Theta\rangle^2}  \label{Omegas}
\end{equation}
and
\begin{equation}
\Omega_{\Lambda}=\frac{3\Lambda}{\langle\Theta\rangle^{2}}\,.
\label{OmegaL}
\end{equation}
We note that $\Omega_{\rho}$, $\Omega_{\sigma}$ and
$\Omega_{\Lambda}$ (assuming $\Lambda>0$) are positive definite,
while $\Omega_{\cal R}$ and $\Omega_{\cal Q}$ can take negative
values. In particular, $\Omega_{\cal Q}$ is negative when ${\cal Q}$
is positive, which happens when the mean-square fluctuation of the
shear is small compared to that of the expansion (see
(\ref{Omegas}d)). In terms of these parameters, and after setting
$\Lambda=0,$ condition (\ref{a3Ricci}) reads
\begin{equation}
\Omega_{\rho}+ \Omega_{\sigma}+ \Omega_{\cal R}+ \Omega_{\cal Q}=
1\,,  \label{a3Ricci2}
\end{equation}
while (\ref{acc}) becomes
\begin{equation}
\Omega_{\cal Q}< -{\textstyle{\frac{1}{4}}}\Omega_{\rho}-
\Omega_{\sigma }\,.  \label{acc1}
\end{equation}
Combining this result with constraint (\ref{a3Ricci2}) we can recast
the condition for accelerated expansion, without any
$\Omega_{\sigma}$ and $\Omega_{\cal Q}$ contributions, into the
following more familiar form~\cite{B1,R2}
\begin{equation}
\Omega_{\rho}+ {\textstyle{\frac{4}{3}}}\Omega_{\cal R}>
{\textstyle{\frac{4}{3}}}\,.  \label{acc2}
\end{equation}
We may constrain the shear anisotropy of the accelerating averaged
domain by combining conditions (\ref{a3Ricci2}) and (\ref{acc2}).
For instance, setting $\Omega_{\cal R}=0$ and substituting
(\ref{acc2}) into (\ref{a3Ricci2}) gives
\begin{equation}
0<\Omega_{\sigma}<-{\textstyle{\frac{1}{3}}}- \Omega_{\cal Q}\,.
\label{ascon}
\end{equation}
This is a self-consistent condition as long as $\Omega_{\cal Q}<
-1/3$, or equivalently ${\cal Q}> 2\langle\Theta\rangle^2/9$ (see
definition (\ref{Omegas}d)). Following (\ref{cQ}), the latter
requires the mean-square fluctuations in the volume expansion to
exceed those in the shear by at least $2\langle\Theta\rangle^2/9$.
Note that, when $\Omega_{\cal Q}\rightarrow-1/3^{-}$, accelerated
expansion occurs when there is minimal shear anisotropy (i.e.~for
$\Omega_{\sigma}\rightarrow0$).

\section{Discussion}\label{sD}
The averaging problem in general relativity and cosmology is an
issue of major importance that remains largely unresolved. As yet
and despite the efforts there is no general consensus on what the
approach to spatial averaging should be. Generally, averaged
inhomogeneous spacetimes are known to evolve differently from the
standard cosmological model, represented by the FRW universe. It is
also known that spatial averaging leads to effective source terms in
the Einstein equations, analogous to Reynolds stress terms in
turbulent fluid flows, which carry what are commonly referred to as
backreaction effects. In fact, these effective dynamical sources are
what an observer would expect to measure. However, although the
measurements take place in the real inhomogeneous world, their
interpretation depends on the chosen fitting cosmological model. So
far, the latter has been the Friedmann universe and the studies have
shown that spatial averaging can seriously modify its standard
dynamics, particularly its expansion rate. In view of the
uncertainty about the averaging procedure and its physical
consequences, it is important to interpret spatial averaging within
a range wider than the FRW models.

In this study we assumed a Bianchi-type cosmology with pressure-free
matter and zero or isotropic 3-curvature. Our intention was to
calculate the effective evolution of the averaged counterparts of
these simple anisotropic models. Allowing for explicit average-shear
terms in the equation, we recast the standard formulae in a closed
Bianchi-type form. Thus, in our case, the backreaction source term
is given by the difference between the mean-square fluctuations in
the expansion and the shear scalars. For a vacuum fitting model with
Euclidean 3-spaces, we found that the source term vanishes and the
averaged equations retain their local form. When pressureless matter
is introduced, however, the backreaction effects can dominate the
early evolution of averaged Bianchi~$I$ models and prevent the
familiar Kasner-type singularity from forming. Assuming axial
symmetry, we have considered the `bounce conditions' for different
types and degrees of anisotropy. The possibility of avoiding the
Kasner regime in averaged inhomogeneous Bianchi~$I$ universes,
raises the question as to whether the same backreaction effects
could also alleviate the Mixmaster behaviour in averaged
Bianchi~$IX$ spacetimes. By giving a Bianchi-type form to the
average equations, we were allowed to consider the limits that a
period of backreaction triggered accelerated expansion might put on
the anisotropy of the averaged domain. Our results show that these
constraints are largely compatible with the observed high isotropy
of the universe.

When averaging relativistic spacetimes one should always keep in
mind that there is no unique approach so far
(e.g.~see~\cite{CPZ}-\cite{CP2}). In addition to that, there is
another central concern that besets all averaging methods and
creates uncertainty about their conclusions (e.g.~see~\cite{IW}).
Averaging takes an ensemble of spacetimes $(M_{i},\,g_{i})$ -- all
of which satisfy the standard Einstein equations -- and introduces a
procedure by which we can define an `average' spacetime
$(\bar{M},\,\bar{g})$ associated with them. This process necessarily
loses information about the $(M_{i},\,g_{i})$ which could lead to
spurious astrophysical conclusions about the behaviour of
$(\bar{M},\,\bar{g})$. For instance, if every member of the ensemble
$(M_{i},\,g_{i})$ is a dust-filled FRW universe with an initial
singularity, then it seems very peculiar if an `average' of these
behaviours, every one of which is singular, could give rise to a
non-singular `average' universe. Yet, this is in effect what is
happening in many of the studies of the averaged behaviour of FRW
universes. Each member of the ensemble obeys the strong-energy
condition and possesses decelerating expansion, but the averaged
universe accelerates and in a sense it effectively violates the
strong energy condition. This is exactly the situation that is being
appealed to in a number of recent studies, and also emerges in our
examples. It remains to be seen whether the extension of the
averaging procedure to deal with non-scalar quantities can produce a
consistent resolution of this information-loss problem by
controlling the type and the quantity of the lost information. We
know that the process of perturbing or averaging the Einstein
equations is not a sequence of operations that commutes. The
definition of an averaged spacetime would be the result of solving
the Einstein equations in general and then averaging the solution.
In practice, we average them and then solve the averaged equations.
This suggests that we might narrow the gap between the two results
of these approaches by averaging an exact inhomogeneous cosmological
solution of Einstein's equations and comparing the result with the
averaging process applied to a FRW ensemble. Our analysis has been
made in this same spirit of testing the self-consistency of an
averaging procedure in new situations to determine whether we can
rely upon the more familiar ones.

\section*{Acknowledgments}
We would like to thank Thomas Buchert, Julien Larena and Aseem
Paranjape for their comments. CGT also wishes to thank the Centre
for Mathematical Sciences at Cambridge University, where part of
this work was conducted, for their hospitality.

\end{document}